\documentstyle[epsf,aps,preprint,rotate]{revtex} 
\tighten 
\begin{document}
\preprint{\small OUTP-97-19P}
\draft

\title{Ruling out a critical density baryonic universe}

\author{\large Michael Birkel and 
 Subir Sarkar~\footnote{PPARC Advanced Fellow\\ 
 \hspace*{2mm} hep-ph/9705331} \bigskip}

\address{Theoretical Physics, University of Oxford, \\ 
          1 Keble Road, Oxford OX1 3NP, UK \bigskip}

\date{submitted to {\sl Physics Letters B}}

\maketitle

\begin{abstract}
It has been suggested by Bartlett and Hall that our universe may have
the critical density in baryons by virtue of specific interactions
with a `shadow' world. We show that this possibility is severely
constrained by primordial nucleosynthesis, stellar evolution and the
thermalization of the cosmic microwave background. In particular,
recent observations of small angular-scale anisotropy in the cosmic
microwave background conclusively rule out all such baryon-dominated
cosmologies. 
\end{abstract}

\bigskip
\pacs{98.80.Cq, 97.60.Bw,98.70.Vc,98.80.Ft}

\small
\section{Introduction}

The `fine tuning' problems of the Friedmann-Robertson-Walker cosmology
favour the possibility that the universe is spatially flat with
density parameter $\Omega=1$, as would result naturally from an
inflationary phase in the early universe \cite{peebles}. However, the
observed abundance of deuterium, which could only have been formed
during big bang nucleosynthesis (BBN) \cite{deut}, implies an upper
bound of $\Omega_{\rm N}<0.033h^{-2}$ on the nucleonic contribution to
the density parameter \cite{etarange}. Since the Hubble parameter
$h~(\equiv\,H_0/100\,{\rm km\,sec}^{-1}{\rm Mpc}^{-1})$ is in the
range $\sim0.5-1$, ordinary matter cannot close the universe and this
discrepancy has motivated the much discussed idea that non-baryonic
dark matter makes up the difference.\footnote{\footnotesize There may
exist new forms of baryonic matter which do not participate in
nucleosynthesis, e.g.\ strange quark matter `nuggets' \cite{witten}. If
these survived baryon evaporation, they could make a substantial
contribution to $\Omega$ today \cite{nuggets}; therefore we
distinguish between baryons and nucleons in the context of BBN.} This
would require the existence of new {\em stable} massive particles in
extensions of physics beyond the Standard Model, e.g. massive
neutrinos or the lightest supersymmetric particle \cite{nbdm}. It is
clearly important to establish whether there are any loopholes in this
argument, especially since nucleons {\em can} in principle make up
all the dynamically detected dark matter which amounts to only
$\Omega\sim0.1$ \cite{peebles,bdm}. In particular, recent observations of
microlensing of stars in the {\sl Large Magellanic Cloud} indicate
that $\sim0.5\,M_{\odot}$ objects (probably old white dwarfs) can
already account for upto half of the dark matter in the halo of our
Galaxy \cite{macho}.

There have been several proposals for relaxing the BBN bound on the
nucleon density by either modifying general relativity \cite{alimi},
or complicating the standard picture of nucleosynthesis, e.g. by
invoking non-zero neutrino chemical potentials \cite{chempot} or
inhomogeneities in the nucleon distribution resulting from a
first-order quark-hadron phase transition \cite{fluc}. The latter
possibility is physically the best motivated but detailed studies show
that even with new free parameters it is difficult to permit the
critical density in nucleons \cite{weirdbbn}. There have also been
suggestions which invoke physics beyond the Standard Model
\cite{bbnrev}. One requires a tau neutrino with a mass of 20--30~MeV
(but with its relic abundance suppressed by a factor of $\sim10$ below
the nominal `freeze-out' value) decaying `invisibly' into electron
neutrinos with a lifetime of 200--1000~sec \cite{nutau}. Although this
would allow a $\Omega_{\rm N}=1$ universe, the model is rather
fine-tuned and quite unlikely from the particle physics viewpoint; in
particular the $\nu_\tau$ must have its tree-level Standard Model
decays ($\nu_\tau\to\,e^+e^-\nu_e$) totally suppressed as these would
disrupt nucleosynthesis \cite{nutaulim,bbnrev}; moreover the required mass
region is now almost ruled out experimentally \cite{pdg}. A better
motivated suggestion \cite{grav} is that the hadronic and radiation
cascades triggered by the decay after BBN of massive unstable
particles such as gravitinos can process the elemental yields so as to
synthesize adequate amounts of deuterium even for $\Omega_{\rm
N}=1$. However, there is a problem with the abundance of $^6$Li, which,
in this model, is nearly as abundant as $^7$Li, in apparent conflict
with observations \cite{Li6}. (The same comment applies to the model
invoking Hawking radiation from primordial black holes
\cite{cascade}.)

In this paper we consider another exotic idea due to Bartlett and Hall
\cite{barthall} viz.\ that our universe (the `visible' sector) may have
become coupled to a `hidden' sector through a phase transition after
nucleosynthesis. The mixing of photons in the visible sector with
their counterpart `paraphotons' in the hidden sector can then result
in a sudden {\em decrease} of the comoving photon density. Thus the
universe may well have the critical density in nucleons today, while
having a nucleon-to-photon ratio during BBN consistent with observed
elemental abundances. To see this we note that the nucleon mass
density in ratio to the critical density is
\begin{equation} 
\label{Omega}
 \Omega_{\rm N} = \frac{m_{\rm N} n_{\rm N}}{\rho_{\rm c}}
                = \frac{8\pi m_{\rm N}}{3 H_0^2 M_{\rm P}^2} 
                   \left(\frac{n_{\rm N}}{n_\gamma}\right)_0
                   \frac{2\zeta(3)T_0^3}{\pi^2}\ , 
\end{equation}
where the present temperature of the blackbody cosmic microwave
background (CMB) is $T_0=2.728\pm0.004$\,K (95$\%$ c.l.)
\cite{mu}. Therefore we require 
\begin{equation}
\label{etacrit}
 \eta_{\rm c} \equiv \left. \frac{n_{\rm N}}{n_\gamma}
                     \right|_{\Omega_{\rm N}=1} \hspace{-6mm}
  \simeq 2.72 \times 10^{-8} h^2 \Theta^{-3}\ ,
\end{equation}
for a critical density in nucleons, where
$\Theta\equiv\,T_0/2.73$\,K. In the standard cosmology, the comoving
entropy does not change after $e^+e^-$ annihilation so the
nucleon-to-photon ratio during nucleosynthesis equals its value today,
i.e. $\eta_{\rm BBN}=\eta_0$. As mentioned earlier, the observational
lower limit to the abundance of deuterium (together with observational
upper limits to the abundances of helium and lithium) requires
$\eta_{\rm BBN}<8.9\times10^{-10}$ \cite{etarange}. The interesting
possibility suggested in ref.\cite{barthall} is that the comoving
photon number {\em decreases} following nucleosynthesis, so that
$\eta_0$ is raised to $\sim(10-100)\,\eta_{\rm BBN}$ and can be as
high as $\eta_{\rm c}$.

We examine the effects of such physics on various cosmological and
astrophysical processes such as BBN itself, the spectrum of the CMB
and stellar evolution. It turns out that these arguments severely
constrain but cannot quite rule out the scenario. However, {\em all}
models in which the critical density is comprised entirely by baryons 
(nucleonic or otherwise) are shown to be definitively excluded by recent
measurements of anisotropy in the CMB together with measurements of
the power spectrum of large-scale structure (LSS) in the universe.
Thus, there appears to be no alternative to non-baryonic dark matter if
indeed $\Omega\gtrsim0.1$.

\section{The photon cooling cosmology} 

First we briefly review the scenario of Bartlett and Hall
\cite{barthall}. They consider a gauge theory with the group
$G{\otimes}G'$ at energies far above the electroweak scale
($v\simeq246$\,GeV). The group $G$ contains the Standard
$SU(3)\otimes\,SU(2)\otimes\,U(1)$ Model while $G'$ corresponds to a
hidden sector with no tree-level renormalizable couplings to known
particles although similar in its symmetry breaking pattern. In
particular $G'$ breaks to a group which includes a hidden (unbroken)
electromagnetism $U(1)'$ at a scale $v'\ll\,v$ and the ${\cal O}(100)$
light degrees of freedom in $G'$ (generically termed $X'$) acquire a
mass at this scale. The cosmology is thus quite novel. The lightest
`connector' particle (carrying both $G$ and $G'$ quantum numbers) has
a mass $M_{\rm C}\gg\,v$ so that scattering processes involving the
connector particles maintain the two sectors in thermal equilibrium at
$T\gg\,M_{\rm C}$. At lower temperatures these reactions freeze-out
and the two sectors decouple. Subsequent phase transitions and
annihilations of heavy species affect the sectors differently
resulting in $T'<T$. (For example if the temperature difference is
dominantly due to heavy particle annihilations in the visible sector
then $T/T'\simeq2\,(11/4)^{1/3}\simeq3$ at $v'\ll\,T\ll\,m_{e}$
\cite{barthall}.) Thus, during BBN the universe contains two plasmas
which are not in thermal contact with each other. At some point after
nucleosynthesis a `recoupling' reaction reestablishes thermal contact,
cooling the photons in the visible sector and hence reducing the
comoving photon number density. Such a reaction requires a
renormalizable interaction between photons and the $X'$ particles in
the hidden sector. Indeed the particles $X'$ can have a very small
electric charge, which may arise naturally in a unified non-Abelian
gauge theory \cite{holdom}, so that the Compton scattering reaction
$\gamma\,X'\to\,\gamma'\,X'$ shown in Figure~1 can recouple the two
sectors.

The rate of this process may be estimated as follows. Wave function
mixing mediated through a one-loop diagram involving the connector
particles generates a term $(\epsilon/2)F^{\mu\nu}F'_{\mu\nu}$ in the
Lagrangian, where $\epsilon\simeq(ee'/16\pi^2)vv'/M_{\rm C}^2$ for
$M_{\rm C}\gg\,v,v'$ and values of $\epsilon\lesssim 10^{-3}$ may be
considered natural \cite{barthall}. After diagonalizing the kinetic
terms, the interactions of the photons can be written in several
alternative bases. Let $J^{\mu}$ and $J'^{\mu}$ be the currents,
including the gauge coupling constants $e$ and $e'$, for U(1) and
U(1)$'$ respectively. When only ordinary matter is relevant it is most
convenient to write the corresponding Lagrangian to order $\epsilon$
in the form
\begin{equation}
\label{lagrange}
 {\cal L}_{\rm int} = a^{\mu} J_{\mu} + (a'^{\mu} + \epsilon a^{\mu}) J'_{\mu},
\end{equation}
where $a^{\mu}$ is defined to be the field coupled to $J^{\mu}$ which
creates ``our'' photon $\gamma$. (Note that the quanta coupled to
$J^{\mu}$ and $J'^{\mu}$ are {\em not} orthogonal. Thus the field
$a'^{\mu}$ does not create the shadow photon $\gamma'$ of the hidden
sector; this is done instead by a field $A'^{\mu}$.) From
eq.(\ref{lagrange}) we see that the hidden particles $X'$ interact
with the photon through a very small electromagnetic charge
$\epsilon\,e'$. For fermionic $X'$ species the rate of the recoupling
process is then \cite{barthall}
\begin{equation}
\label{rate}
 n_{X'} \langle\sigma\,v\rangle = 
  C\,n_{X'}\,\frac{\pi\alpha'^2\epsilon^2}{T\,T'}
\end{equation}
where $\alpha'=e'^2/4\pi$ and $C$ is a number of ${\cal O}(1)$ which
comes from thermally averaging over the energy transfer. As we shall
see, constraints from the CMB require the recoupling reaction to occur
during the radiation-dominated era. The energy density of the
universe is then dominated by the plasma in the visible
sector. Defining the recoupling temperature $T_{\rm R}$ to be the
temperature of our plasma when the recoupling rate equals the 
expansion rate gives \cite{barthall}
\begin{equation}
\label{alphatilde}
 \tilde\alpha \simeq 10^{-22} \frac{1}{C} \left(\frac{T}{T'}\right)^2_{\rm R}
   \left(\frac{T_{\rm R}}{10 \mbox{ keV}}\right) \left(\frac{0.1}{\alpha'} 
   \right)\ ,
\end{equation}
where $\tilde\alpha\equiv\epsilon^2\alpha'\sum_{i}g'_i\,q^2_i$ and
$g'_i$ is the number of spin states (assumed fermionic) of the hidden
plasma having interaction $q_i\,e'$ with $\gamma'$. To calculate the
resultant cooling of the visible sector it is assumed that the energy
transfer occurs quickly compared with the expansion rate; the final
temperature $T_{\rm F}$ is then obtained by demanding energy
conservation at recoupling, i.e. $gT^4_{\rm R}+g'T'^4_{\rm
R}=(g+g')T^4_{\rm F}$. The energy density of the shadow sector should
be negligible during BBN (in order not to increase the expansion rate
excessively) but dominates after recoupling (because $g'\gg\,g=2$) which
gives $n_\gamma(T_{\rm F})/n_\gamma(T_{\rm
R})\simeq(g/g')^{3/4}$. Therefore a critical density
nucleonic universe (see eq.(\ref{etacrit})) requires $g'\sim40-1000$
and, correspondingly, $T/T'\gtrsim2-5$. Subsequently the $X'$
particles acquire mass through a phase transition with a critical
temperature bounded as
\begin{equation}
\label{phasetran}
 T'_{\rm C} < T'_{\rm R} < T_{\rm R}/2\ .
\end{equation}  
It would be natural to expect that $m_{X'}$ is of ${\cal O}(T'_{\rm
C})$.

\subsection{Constraint from BBN}

Nucleosynthesis obviously provides an upper bound on the recoupling
temperature since the abundances of the elements would be altered if
it is too high. To study this, we have modified the standard computer
code \cite{bbncode} to include an abrupt drop in temperature at
$T_{\rm R}$ when the recoupling reaction occurs. Before this, the
universe contains two plasmas which are not in thermal contact, the
usual one characterized by the photon temperature $T\equiv\,T_\gamma$
(consisting of $\gamma$, 3 $\nu$'s, $e^\pm$, as well as nucleons), and
the other one containing the shadow sector particles like $\gamma'$
and $X'$ with temperature $T'$. An important parameter is the ratio of
the shadow sector temperature to the neutrino temperature {\em before}
recoupling,
\begin{equation}
 r \equiv \frac{T'}{T_\nu}\ ,
\end{equation} 
which stays constant during $e^+e^-$ annihilation when the photon
temperature $T$ increases (but not $T'$ or $T_\nu$). It enables the
energy density $\rho'$ of the shadow sector before recoupling to be
tracked using the neutrino energy density,
$\rho'=\rho_\nu\,g'r^4/g_\nu$, and is thus a more convenient parameter
than $T'/T$ as used in ref.\cite{barthall}. Now at the recoupling
temperature $T_{\rm R}$ there is a sudden drop (which would be smooth
in a more detailed model) to the final temperature $T_{\rm F}$. Energy
conservation gives
\begin{equation}
 \case{\pi^2}{30} (g_\gamma + g_\nu + g_{\rm e}) T_{\rm R}^4 + 
  \rho'(T'_{\rm R}) 
 = \case{\pi^2}{30} (g_\gamma + g_\nu + g_{\rm e})
                  T_{\rm F}^4 + \case{\pi^2}{30} g' T_{\rm F}^4 \ , 
\end{equation}
where $g_\gamma=2$, $g_\nu=21/4$ and $g_{\rm e}=7/2$. Since $T_{\rm
R}=T_{\rm F}$ for neutrinos, the corresponding terms above cancel and
we obtain (neglecting the electron energy density for the temperatures
of relevance),
\begin{equation}
\label{tfin}
  T_{\rm F} = A\,T_{\rm R}\ , \qquad
  A \equiv \left[\frac{2 + (\case{4}{11})^{4/3} 
        g' r^4}{2+g'}\right]^{1/4}\, .
\end{equation}
After recoupling the energy density of the shadow sector is just
$\rho'=\case{\pi^2}{30}g'T^4$ since now $T=T'$.

We have implemented a routine in the BBN code dividing the thermal history
into the three steps (before, during and after recoupling) and compute
the expansion rate accordingly, taking into account the contribution
of shadow particles. We assume 3 light neutrino species, use the
neutron lifetime $\tau_{n}=887\pm2$~s \cite{pdg}, and incorporate
small corrections to the helium abundance as reviewed in
ref.\cite{bbnrev}. Since the main effect of photon cooling is to
increase the helium-4 abundance and decrease the deuterium abundance,
we bound their values from recent observations of helium in metal-poor
extragalactic HII regions \cite{izotov} and of deuterium in the
interstellar medium \cite{deutism}:
\begin{equation} 
\label{abund}
 Y_{\rm P}(^4{\rm He}) \leq 0.25 , \qquad
 ({\rm D/H})_{\rm p} \geq 1.1 \times 10^{-5} .
\end{equation}
We do not consider the bound on $^7{\rm Li}$ because it does not
provide a useful constraint. The adopted bounds are very {\em
conservative}, based upon consideration of a variety of data which are
critically discussed elsewhere \cite{bbnrev}. As shown in Figure~2, we
can then derive the maximum value of $T_{\rm R}$ allowed for each
given set of parameters $r$ and $g'$. The allowed region is:
\begin{equation}
\label{bbnresult}
  T_{\rm R} < 74 \mbox{ keV}\ , \qquad
  r < 0.3\  \qquad \mbox{and} \qquad
  40 < g' < 2340\ ,
\end{equation}
only part of which is shown in the figure for convenience. (The
highest allowed value of $g'$ corresponds to the smallest initial
value of $\eta$, viz.\ $\eta_{\rm BBN}>1.7\times10^{-10}$
\cite{etarange}.) For comparison, the values suggested in
ref.\cite{barthall} were $T_{\rm R}\lesssim\,30$~keV, $r\lesssim\,0.7$
and $g'\sim40-1000$.

\subsection{Constraint from the CMB spectrum}

A lower bound on the recoupling temperature follows from the
observation by COBE that the CMB spectrum is accurately blackbody
\cite{mu}. The recoupling reaction {\em removes} photons from the
thermal bath thus inducing a chemical potential for the photon
distribution. If the reaction occurs early enough, however, photon
creation processes can restore the Planck spectrum \cite{husilk} in
agreement with the observational upper limit \cite{mu} on such a
chemical potential: $\mu<9\times10^{-5}$ (95$\%$ c.l.). To thus
constrain $T_{\rm R}$ we use a simple argument concerning the relevant
time scales.

At temperatures below the $e^+e^-$ threshold, the processes which
determine the spectral shape are Compton scattering (C) which
preserves the photon number, and the radiative processes
bremsstrahlung (B) and double Compton scattering (DC). The time scales
of these processes are \footnote{\footnotesize We have inserted the correction
factor of $4\pi$ \cite{husilk} into the rate for bremsstrahlung
\cite{tscales}.}
\begin{eqnarray}
\label{scales}
 t_{\rm C}  & = & 9.25 \times 10^{-7} \,{\rm sec} \ \frac{1}{\eta} 
           \left(\frac{T}{\rm keV}\right)^{-4}, \nonumber \\
 t_{\rm DC} & = & 2.53 \times 10^{-2} \,{\rm sec} \ \frac{1}{\eta} 
           \left(\frac{T}{\rm keV}\right)^{-5}, \nonumber \\
 t_{\rm B}  & = & 2.18 \times 10^{-8} \,{\rm sec} \ \frac{1}{\eta^2} 
           \left(\frac{T}{\rm keV}\right)^{-5/2} \ln^{-2}
           \left[2.13 \frac{1}{\sqrt{\eta}} 
           \left(\frac{T}{\rm keV}\right)^{3/4}\right].
\end{eqnarray}
For a $\Omega_{\rm N}=1$ universe, the value of $\eta$ after
recoupling (i.e. today) is given by eq.(\ref{etacrit}) whereas before
recoupling it is
\begin{equation}
 \eta (T \ge T_{\rm R}) 
 = \eta_0 \,\frac{n_\gamma (T_{\rm F})}{n_\gamma (T_{\rm R})} 
 = \eta_0 \,A^3 \ ,
\end{equation}
where $A$ is defined in eq.(\ref{tfin}). (Note that baryons and
nucleons are equivalent with regard to the CMB since only their
electromagnetic interactions are relevant; therefore we henceforth 
refer to $\Omega_{\rm B}$ rather than $\Omega_{\rm N}$.) The time
scales above are to be compared with the age of the universe obtained
by integrating the Friedmann equation,
\begin{equation}
 t = \left\{  
  \begin{array}{lll}
   2.42 \times 10^6 \,{\rm sec} \, g_{\rm i}^{-1/2} \left( 
   \frac{T}{\rm keV} \right)^{-2} & \mbox{for} & T > T_{\rm R}\ , \\
   2.42 \times 10^6 \,{\rm sec} \, \left[\left(\frac{1}{\sqrt{g_{\rm i}}} -
   \frac{1}{A^2 \sqrt{g_{\rm F}}}\right) 
   \left(\frac{T_{\rm R}}{\rm keV}\right)^{-2}
   + \frac{1}{\sqrt{g_{\rm F}}} \left( 
   \frac{T}{\rm keV} \right)^{-2} \right] & \mbox{for} & T \le T_{\rm F}\ .
  \end{array}                
  \right.
\end{equation}
where we have used
\begin{eqnarray}
 g_{\rm i} &\equiv g (T > T_{\rm R}) &= 2 + \case{21}{4}
  \left(\case{4}{11}\right)^{4/3} + g' r^4
  \left(\case{4}{11}\right)^{4/3}, \\ \nonumber
 g_{\rm F} &\equiv g (T \le T_{\rm F}) &= 2 + \case{21}{4} 
  A^{-4} \left(\case{4}{11}\right)^{4/3} + g' \ .
\end{eqnarray}
Obviously the abrupt transition at $T_{\rm R}$ would be smoothed out
in a more detailed picture of the phase transition. 

In Figure~\ref{fig3} we show the various time scales for a typical set
of parameters, with a sudden jump in $\eta$ as the temperature drops
from $T_{\rm R}$ to $T_{\rm F}$ as determined by
eq.(\ref{tfin}). Although bremsstrahlung usually dominates at low
redshifts for a high density baryonic universe, the most efficient
photon creation process is always found to be double Compton
scattering for the values of $T_{\rm R}$, $r$ and $g'$ under
consideration. The limiting value of the recoupling temperature is
then conservatively obtained by demanding that the time scale for this
be shorter than the expansion time scale. We have indicated in the
figure the error bands for the time scales in eq.(\ref{scales})
corresponding to the allowed region of $\eta$ after nucleosynthesis,
viz.\ $\eta\in(1.7-8.9)\times10^{-10}$ \cite{etarange}. To obtain the
minimum recoupling temperature we took the highest possible value for
$\eta_0$ from eq.(\ref{etacrit}) and searched for the temperature at
which $t\ge\,t_{\rm DC}$ is satisfied as a function of the parameters
$r$ and $g'$; this gives
\begin{equation}
\label{trmin}
 T_{\rm R} \ge 0.73 \,{\rm keV}\ \left(\case{\sqrt{g_{\rm 
                                  i}}}{A^5}\right)^{1/3}\ . 
\end{equation} 
Combining the constraints from BBN and the CMB spectrum it follows
that the recoupling temperature must be in the range
\begin{equation}
\label{bbnCMB}
 3 \,{\rm keV} <  T_{\rm R} < 75 \,{\rm keV}  
\end{equation}
for the allowed range of $r$ and $g'$ as given in
eq.(\ref{bbnresult}).

\subsection{Constraints from stellar evolution}

Further constraints on the model follow from the energy loss argument
for stars \cite{barthall}. For example if the $X'$ particles are light
enough to be produced in red giants through plasmon decays, a limit on
$\tilde\alpha$ (eq.(\ref{alphatilde})) follows from requiring that the
energy loss through $X'$ emission does not affect the stellar
luminosity or age to a noticeable degree \cite{radesil}. The resultant
constraint is very stringent, viz.\ $\tilde\alpha\lesssim10^{-28}$
\cite{barthall}, and in conflict with the required value from
eq.(\ref{alphatilde}) unless the recoupling temperature is very low. This,
however, is not allowed by the CMB spectrum constraint (eq.(\ref{trmin})) so
the only escape route is to argue that the $X'$ particles are too
heavy to be produced in stars \cite{barthall}. In the cores of red
giants the plasmon mass is $\omega_{\rm pl}\sim20$~keV, so this
requires $m_{X'}>\omega_{\rm pl}/2\gtrsim10$~keV. There are additional
contributions to $X'$ production from Compton and bremsstrahlung
processes and to suppress these one would require $m_{X'}\gtrsim\,5T$
where $T\sim10$~keV is the core temperature
\cite{barthall}.\footnote{\footnotesize Note that processes of the
form $e^-\gamma\to\,e^-\gamma'$ or
$e^-\,^4\mbox{He}\to\,e^-\,^4\mbox{He}\gamma'$ are not allowed by the
Lagrangian of eq.(\ref{lagrange}); if these existed, the associated
energy loss could {\em not} be evaded by increasing $m_{X'}$.} However,
if $m_{X'}$ is indeed of ${\cal O}(T_{\rm R})$ then this implies
$T_{\rm R}\gtrsim50$~keV (see eq.(\ref{phasetran})), which exceeds the
upper limit from BBN over most of the parameter space (see
Figure~\ref{fig2}). Thus it would be necessary to invoke some
fine-tuning and argue that $m_{X'}\gg\,T_{\rm R}$.

An even more restrictive constraint on $\tilde\alpha$ comes from
supernova physics. After collapse, the supernova core forms a neutron
star which radiates its binding energy primarily in the form of
neutrinos which diffuse out of the dense central region over a time
scale of seconds. Observations of (anti)neutrinos from {\sl SN1987A}
\cite{kamio} confirmed that these indeed carried away $\sim10^{53}$
erg on the expected time scale. A competing mechanism for cooling the
core would be the production of $X'$ particles through plasmon decays.
Then a conservative requirement is that the $X'$ particles do not carry
away more energy than the neutrinos and thus curtail the duration and
energetics of the neutrino burst. Assuming the core radius to be
$\sim10$~km and a cooling time of $\sim10$~s one obtains for the
energy loss $E'$ through $X'$ \cite{barthall}
\begin{equation}
\label{snengy}
 E' \approx 10^{71}\ {\rm erg}\ \tilde\alpha \rho_{14}^{2/3} T^3_{10} 
    \lesssim 10^{53}\ {\rm erg}\ ,
\end{equation}
where $T_{10}\equiv\,T/(10~\mbox{MeV})$,
$\rho_{14}\equiv\rho/(10^{14}~{\rm g}/{\rm cm}^3)$, and the $X'$
particles are assumed to be lighter than the plasmon mass in the
supernova core. Typical values for the density and temperature are
$\rho_{14}\sim6$ and $T_{10}\sim5$ \cite{raffelt}. Solving this
equation for $\tilde\alpha$ and inserting into eq.(\ref{alphatilde})
we then find:
\begin{equation}
\label{sntr}
 T_{\rm R} \lesssim 3.6 \times 10^5\ {\rm keV}\ \rho_{14}^{-2/3}
                   T_{10}^{-3} C \alpha'\, r^2\ .
\end{equation}
In Figure~\ref{fig4} we show that this constraint together with those
from BBN and the CMB spectrum effectively rules out the model (taking
$C=1$, and $\alpha'=0.1$). The only allowed region is a tiny corner at
the highest value of $r$ and the lowest value of $g'$, fixing the
recoupling temperature to be 7--8~keV. This region is further limited
by decreasing $\alpha'$ or $C$. Of course the physical parameters of
supernova collapse are somewhat uncertain so this argument is not
definitive. One can only conclude that $X'$ particles, if they exist,
may play an essential role in the supernova process \cite{barthall}.

\subsection{Constraints from LSS and CMB anisotropy}

Finally we consider the formation of large scale structure and the
concomitant generation of small scale CMB anisotropy in a critical
density baryonic universe.\footnote{\footnotesize Note that in the
Bartlett-Hall model \cite{barthall} the $X'$ particles of the shadow
sector can be a candidate for the dark matter. Since
$n_{X'}\sim\,n_\gamma$ at recoupling, there will be a large
contribution to the density parameter
($\Omega_{X'}\sim40h^{-2}(m_{X'}/{\rm keV})$) unless the $X'$ particles
subsequently self-annihilate efficiently after becoming
non-relativistic. This is however quite possible and, in fact, very
likely in analogy to the `visible' sector.} In the standard picture,
such structure grows by gravitational instability from a (nearly)
scale-invariant spectrum of primordial adiabatic, gaussian
fluctuations (with power spectrum $P(k)\propto\,k^n$, $n\approx1$),
presumably generated during an inflationary phase in the early
universe \cite{structure}. On scales smaller than the horizon size at
matter-radiation equality the primordial spectrum is modified and its
slope becomes negative. The COBE detection of anisotropy in the CMB on
angular scales exceeding the horizon size at the recombination epoch
\cite{cobe} has provided strong support for this basic picture and
moreover fixed the normalization of the primordial
spectrum. Subsequent observations of anisotropy on small angular
scales have also provided evidence for a `Doppler peak' in the angular
power spectrum such as would be expected from acoustic oscillations of
the coupled plasma and photon fluids during the last scattering of the
CMB \cite{structure}. Besides the observed power spectrum of galaxy
clustering, we thus have a new quantitative test for models of
structure formation.

A baryonic universe was of course the first to be investigated in the
context of structure formation \cite{nuclss}. The salient feature of
such a universe is that fluctuations can begin to grow only after
(re)combination (rather than as soon as the universe becomes
matter-dominated) since the baryons are tightly coupled to the photons
before this epoch. Moreover the coupling is not perfect (and becomes
weaker as the universe turns neutral) so adiabatic fluctuations suffer
`Silk damping' on scales smaller than the damping mass of $M_{\rm
D}\approx1.3\times10^{12}(\Omega_{\rm B}h^2)^{-3/2}M_\odot$. It was
recognized early on that primordial gaussian fluctuations with
spectral index $n\lesssim2$ would thus result in excessively large
fluctuations in the mass distribution on small scales
\cite{peeb}. Moreover the upper bounds on CMB anisotropy available a
decade ago already implied that there would not be sufficient time
after (re)combination for the implied small primordial fluctuations to
grow to the large-scale structure observed today in a baryonic
universe \cite{bondefst}.

In the past particle physicists have tended to ignore such constraints
since there did not exist a `Standard Model' for the formation of
structure. The increasing weight of observational evidence for the
picture outlined above, however, now makes such a viewpoint
untenable. In particular the data (as opposed to upper limits) on CMB
anisotropy now allow a precise quantitative test of a baryon-dominated
universe. For a $\Omega_{\rm B}=1$ universe, the slope of the
primordial spectrum $n$ and the Hubble parameter $h$ are the only free
parameters. Thus we will first determine the values of $n$ and $h$
which would allow a fit to the data on small-scale CMB
anisotropy. These are available as measurements of the multipole
moments $C_l=\frac{1}{2l+1}\sum_{m} a_{lm}^2$, expanding the CMB 
temperature on the
celestial sphere in spherical harmonics:
\begin{equation}
  \Delta\,T = \sum_{l,m} a_{lm} Y_{lm} (\theta, \phi)\ .
\end{equation}
Then we can check if the power spectrum of galaxy clustering can be
reproduced for the {\em same} choice of parameters. We emphasize that
this test is thus as model-independent as possible.

To compute the angular power spectrum for each model specified by a
pair $(n,h)$, we use a fast Boltzmann code \cite{selzal} and adopt the
COBE-normalization at large angular scales. Then we convolve each
power spectrum with the experimental `window functions' before
comparison with the observational data by generating a $\chi^2$
surface over the $(n,h)$ space under investigation; this method is
outlined in detail in ref.\cite{linewea}. The $\chi^2$ value is
computed according to
\begin{equation}
 \chi^2 (n, h) = \sum_{N=1}^{N_{\rm exp}} \left[\frac{\delta 
   T^{\rm data}_{l_{\rm eff}}(N) -
   \delta T^{\rm model}_{l_{\rm eff}}(N,n,h)}{\sigma^{\rm data}(N)} \right]^2, 
\end{equation}
where the sum is over all the CMB observations, $\delta\,T^{\rm
model}_{l_{\rm eff}}(N,n,h)$ and $\delta\,T^{\rm data}_{l_{\rm
eff}}(N)$ are, respectively, the flat band powers (i.e. convolved with
the experiment-specific `window functions') for the models and
observations, and $\sigma^{\rm data}(N)$ are the associated
observational errors. The CMB data we consider and the corresponding
errors are tabulated in \cite{linewea} and shown in
Figure~\ref{fig5}. (We consider only the central
value of the Saskatoon data which have a $\pm14\%$ calibration
uncertainty.) The band powers are given by
\begin{equation}
\label{deltaTmod}
 \delta T_{l_{\rm eff}}^{\rm model} = \left[\frac{1}{I(W_{l}(N))} 
  \sum_{l=2}^{l_{\rm max}} \frac{2l+1}{4\pi} C_{l}(n,h) W_{l}(N) \right]^{1/2},
\end{equation}
where the deconvolving factors $I(W_l(N))$ are the logarithmic
integrals of the experimental window functions $W_l(N)$ defined as
\begin{equation}
 I(W_l) = 2 \pi \sum_{l=2}^{l_{\rm max}} \frac{(2l+1)W_l}{4 \pi l(l+1)}, 
\end{equation} 
and $l_{\rm max}$ can be safely taken to be 1200 for all existing data
(for larger $l$ the acceptance is negligible). Finally, the angular
scale investigated by experiment $N$ is given according to
\begin{equation}
 l_{\rm eff}(N) = \frac{I(lW_l(N))}{I(W_l(N))}\ .
\end{equation}
To obtain a $\chi^2$ value for each point of the $(n,h)$ space we use
the data and eq.(\ref{deltaTmod}) for each observation $N$ and sum
over all existing observations. The values of the multipoles
$C_{l}(n,h)$ are obtained by running the fast Boltzmann code mentioned
above.\footnote{\footnotesize We ignore a possible contribution from
gravitational waves (which would lower the COBE normalization) and
adopt for the helium fraction the central value $Y_{\rm P}(^4{\rm
He})=0.24$ \cite{izotov}. Note also that the error bars
$\sigma^{\rm data}(N)$ are mostly asymmetric; to accommodate this in the
$\chi^2$ calculation, we establish for each data point whether the
model lies above or below and then take the appropriate error bar in
eq.(\ref{deltaTmod}).}

In Figure~\ref{fig6} we show the region of $(n,h)$ space which is
consistent with the data. We investigate the parameter range
$0.25\leq\,h\leq1$ and $0.5\leq\,n\leq1.5$, the former covering {\em
all} direct measurements \cite{pdg} and the latter based on general
theoretical considerations of inflationary models \cite{lyth}. We show
both the $68\%$ C.L. and $95\%$ C.L. contours; the latter delineate
the range
\begin{equation}
\label{nhreslts}
 0.59 \leq n \leq 0.78\ , \qquad 0.41 \leq h \leq 0.62\ .
\end{equation}
The negative correlation of $n$ and $h$ arises because the amplitude
of the acoustic peak plays the dominant role in the fitting procedure
and for $\Omega_{\rm B}=1$ this peak amplitude decreases as $h$ and
$n$ decrease. Thus a decrease in $n$ accompanied by a correlated
increase in $h$ (or vice versa) can preserve the amplitude and thus
the $\chi^2$ value for the fit. In general rather low values of both
$n$ and $h$ are required to fit the acoustic peak in a critical
density baryonic universe.

Given these favoured values of $n$ and $h$ we can now calculate the
linear matter power spectrum. In Figure~\ref{fig7} we show the spectra
for the two extreme models in the $(n,h)$ plane, viz.\ $n=0.5$,
$h=0.62$ and $n=0.78$, $h=0.25$, along with the data points inferred
from the APM galaxy survey \cite{apm}. To compute these spectra we have
used the COSMICS code \cite{bertsch} and normalized to COBE using the
prescription given in ref.\cite{cobenorm}. The strong oscillations
seen at short scales arise from matter fluctuations on wavelengths
smaller than the Hubble radius at (re)combination and reflect the
value of the phase of the perturbation at recombination
\cite{peeb}. It is seen that even for extreme choices of parameters,
the power spectra of $\Omega_{\rm B}=1$ models disagree strongly with
the observational data. Even though the clustering of APM galaxies may
be enhanced over that of the matter distribution, the discrepancy
cannot be resolved for any sensible value of the `bias' parameter. The
situation is further aggravated if the value of $h$ is decreased in
Figure~\ref{fig7}(a) or the value of $n$ is decreased in
Figure~\ref{fig7}(b), staying within the allowed region of
eq.(\ref{nhreslts}), since in both cases the matter power spectrum is
further damped at short scales. Thus we can firmly establish that it
is not possible to simultaneously satisfy the LSS and CMB data in an
universe with the critical density in baryons.

\section{Discussion}

Although big bang nucleosynthesis is usually cited as providing the
main argument for non-baryonic dark matter, it is possible to
entertain deviations from the standard picture in order to accommodate
a baryon-dominated critical density universe. One such suggestion
invoking a {\em reduction} in the comoving entropy after
nucleosynthesis \cite{barthall} is particularly interesting because it
challenges the standard assumption that the thermodynamic history of
the universe is well known at least up until the BBN epoch. We have
demonstrated that this model is very tightly constrained (if not ruled
out altogether) from consideration of other processes such as
thermalization of the CMB spectrum and the effects on stellar
evolution of the new physics invoked by the model. Finally we have
shown that recent observations of small-scale CMB anisotropy and
large-scale structure decisively rule out this model. This conclusion
also holds for other models which attempt to evade the constraint from
nucleosynthesis in order to allow a baryonic Einstein-De Sitter
universe.

\vfill \noindent {\bf Acknowledgments.} We would like to thank
Charles Lineweaver for supplying the window functions for CMB
experiments, Uros Seljak and Matias Zaldarriaga for providing the
Boltzmann code, and Leith Cooper, Lawrence Hall, Georg Raffelt 
and Matias Zaldarriaga for helpful discussions. M.B. gratefully 
acknowledges financial support from the
Fellowship HSP II/AUFE of the German Academic Exchange Service
(DAAD). This work was supported by the EC Theoretical Astroparticle
Network CHRX-CT93-0120 (DG12 COMA).

\begin{figure}[t]
\center{\epsfysize7cm\epsffile{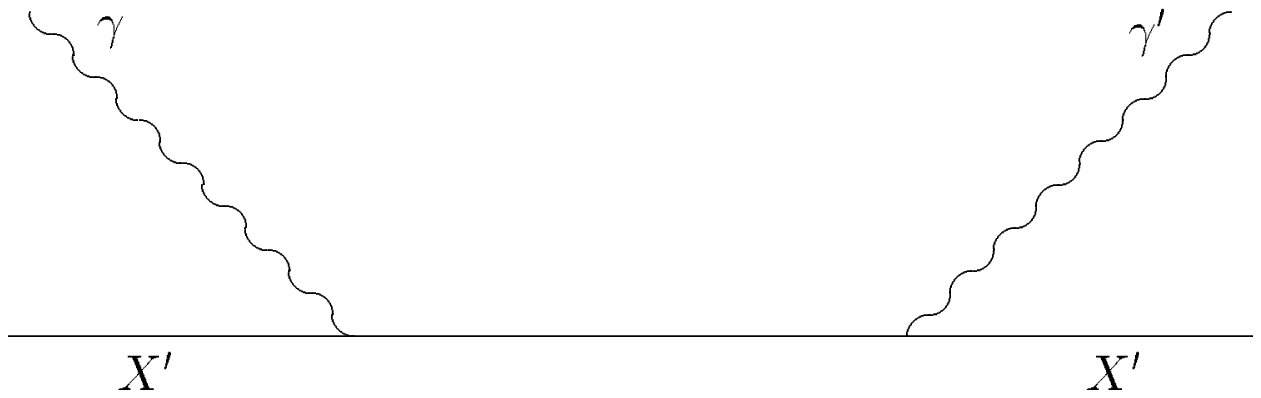}}
\caption{The Bartlett-Hall mechanism for recoupling of the two sectors
via photon mixing.}
\label{fig1}
\end{figure} 
\begin{figure}[t]
\epsfxsize\hsize\epsffile{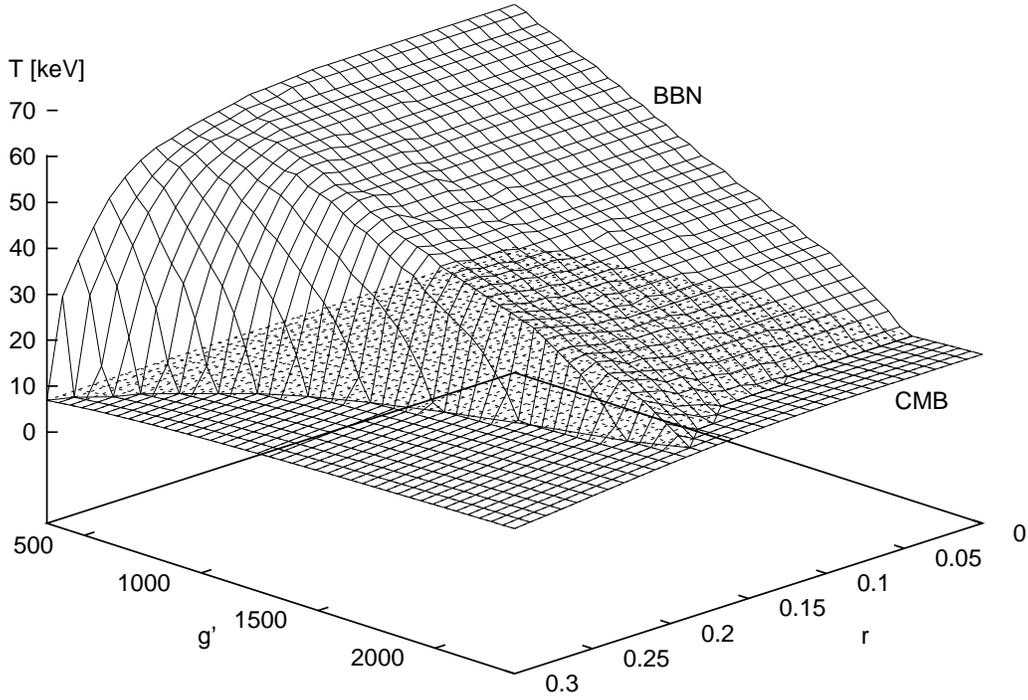}
\caption{Constraints on the recoupling temperature as a function of
 the model parameters $r$ and $g'$ from BBN and the CMB spectrum. The
 {\em allowed} region is between the two surfaces.}
\label{fig2}
\end{figure} 
\begin{figure}[t]
\center{\epsfxsize\hsize\epsffile{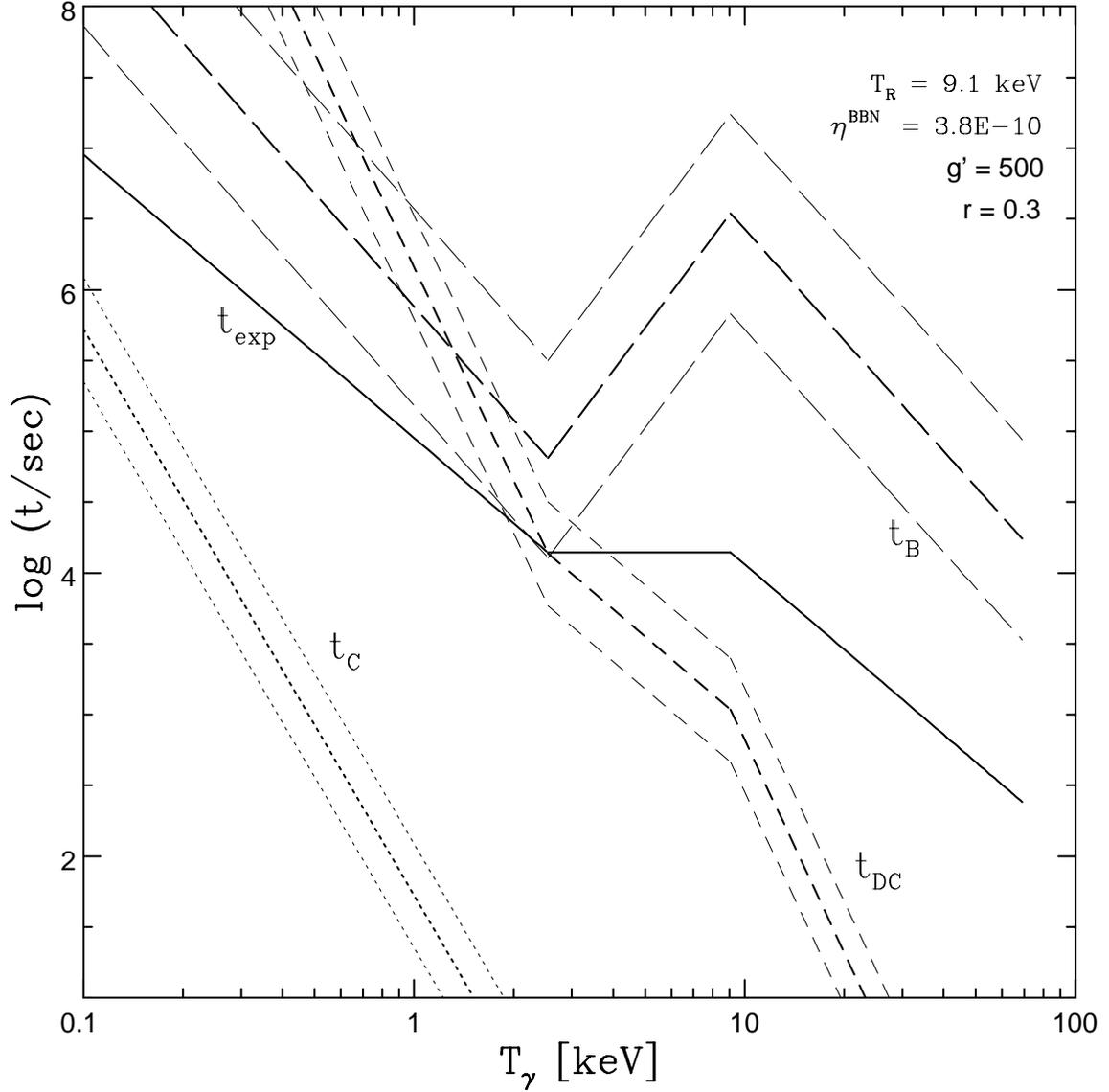}}
\caption{Time scales for Compton scattering, double Compton scattering
 and bremsstrahlung compared with the expansion time scale in the
 Bartlett-Hall model. The steps correspond to the
 recoupling process at which the temperature drops from $T_{\rm R}$ to
 $T_{\rm F}$ and the comoving photon density decreases by a factor of
 $\sim10 - 100$.}
\label{fig3}
\end{figure} 
\begin{figure}[t]
\epsfxsize14cm\epsffile{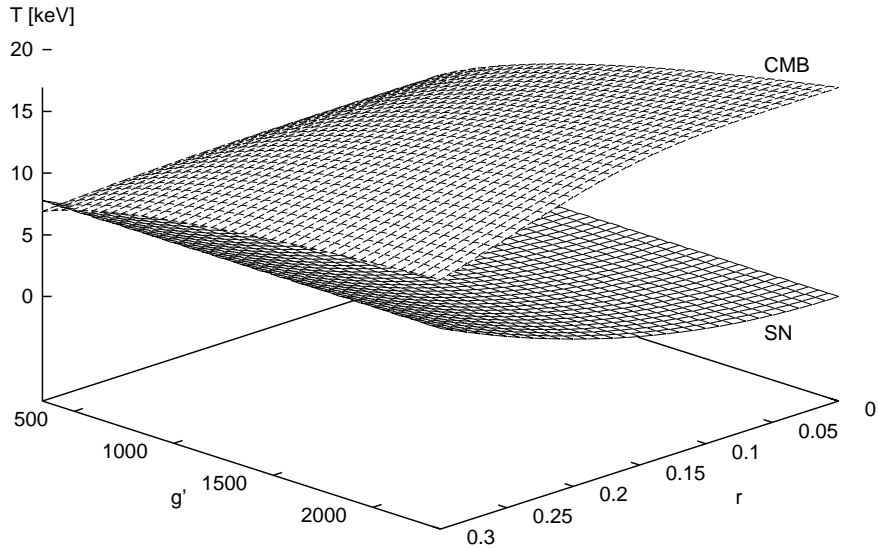}
\caption{Constraints on the recoupling temperature as a function of
 the model parameters $r$ and $g'$ from the CMB spectrum and supernova
 physics (for $\rho_{14}=6$, $T_{10}=5$, $C=1$ and $\alpha'=0.1$). The
 region between the surfaces is {\em excluded}. The only allowed
 region obtains for the highest values of $r$ and the lowest for $g'$,
 fixing the recoupling temperature to be $\sim7$ keV.}
\label{fig4}
\end{figure} 
\begin{figure}[t]
\center{\epsfxsize13cm\epsffile{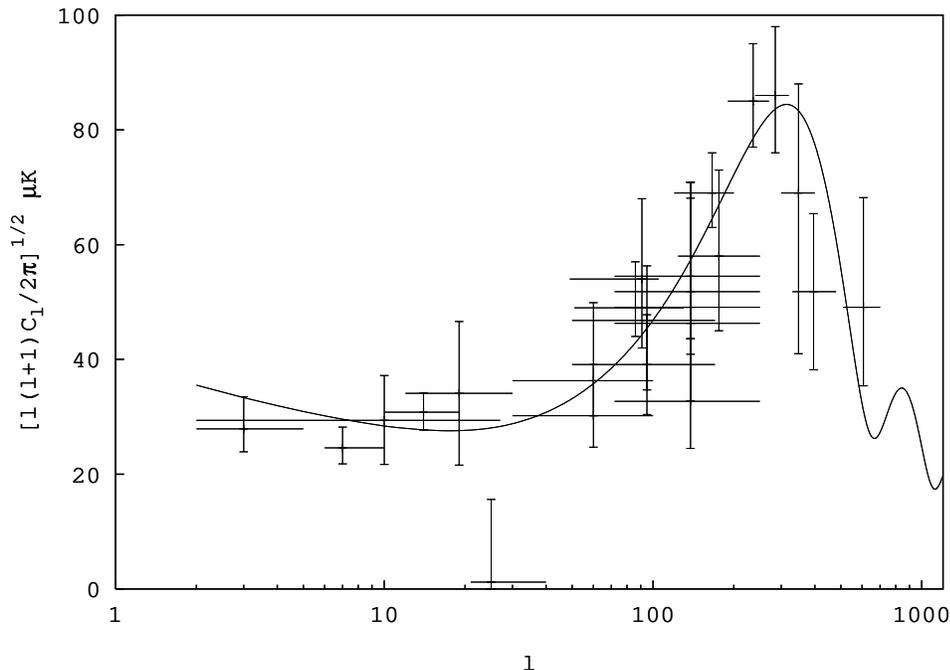}}
\caption{An illustrative fit to the data on the CMB angular power
 spectrum of a $\Omega_{\rm B}=1$ universe with primordial spectrum
 slope $n=0.66$ and Hubble parameter $h=0.3$).}
\label{fig5}
\end{figure} 
\begin{figure}[t]
\center{\epsfxsize\hsize\epsffile{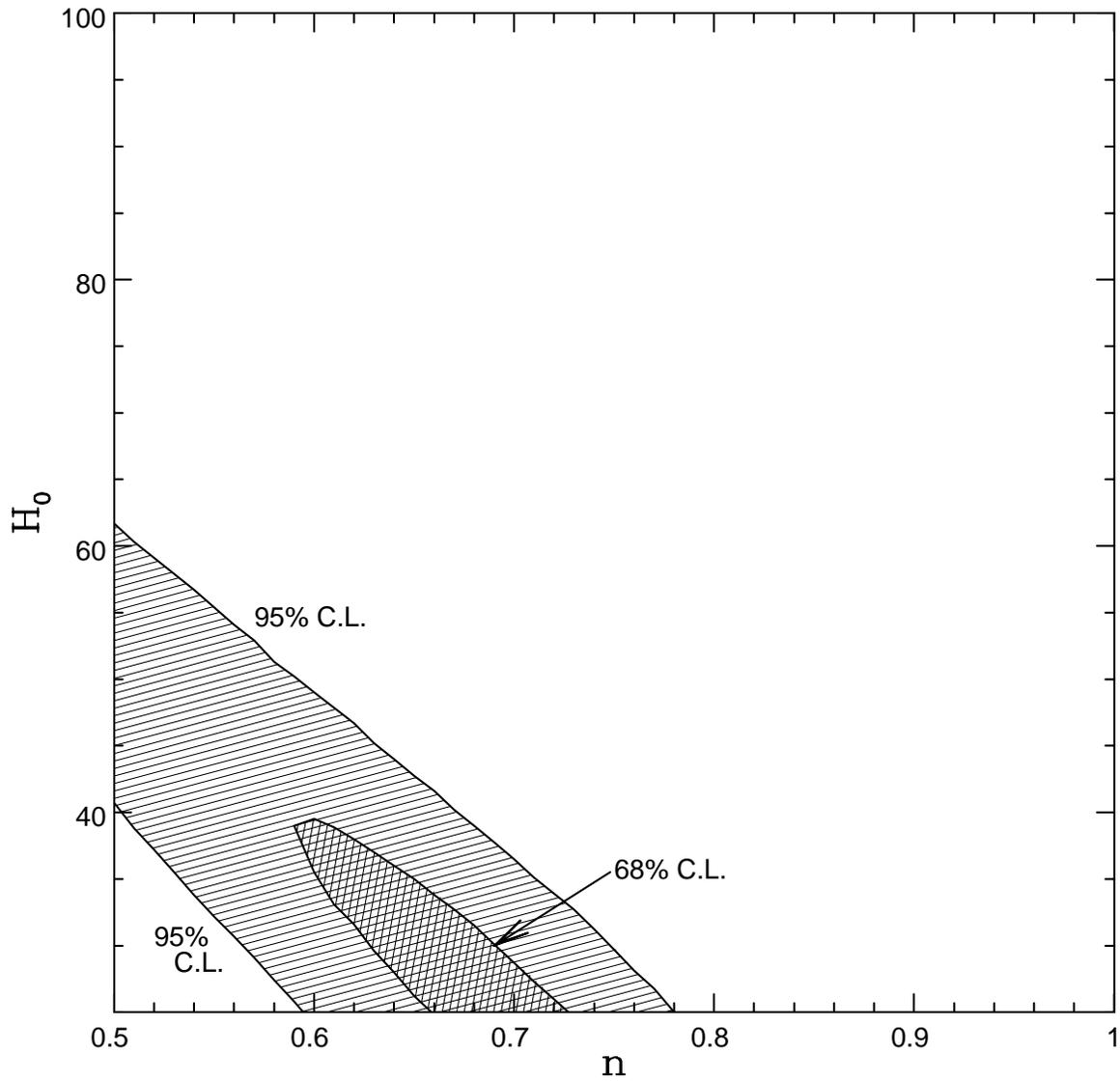}}
\caption{$\chi^2$ contours (68\% and 95\% C.L.) in the $n-H_0$ plane
 for the fit to observations of CMB anisotropy of a $\Omega_{\rm B}=1$
 universe. The negative correlation between the primordial spectral
 index and the Hubble parameter is discussed in the text.}
\label{fig6}
\end{figure} 
\begin{figure}[t]
\center{\epsfxsize15cm\epsffile{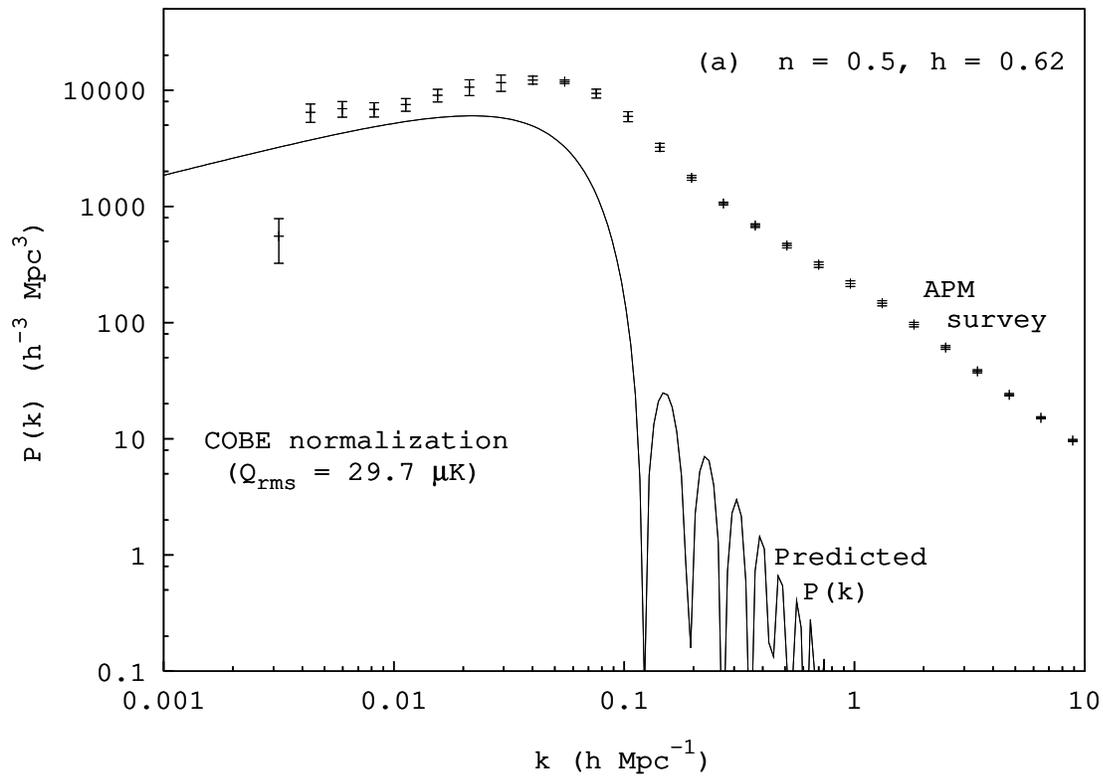}
\epsfxsize15cm\epsffile{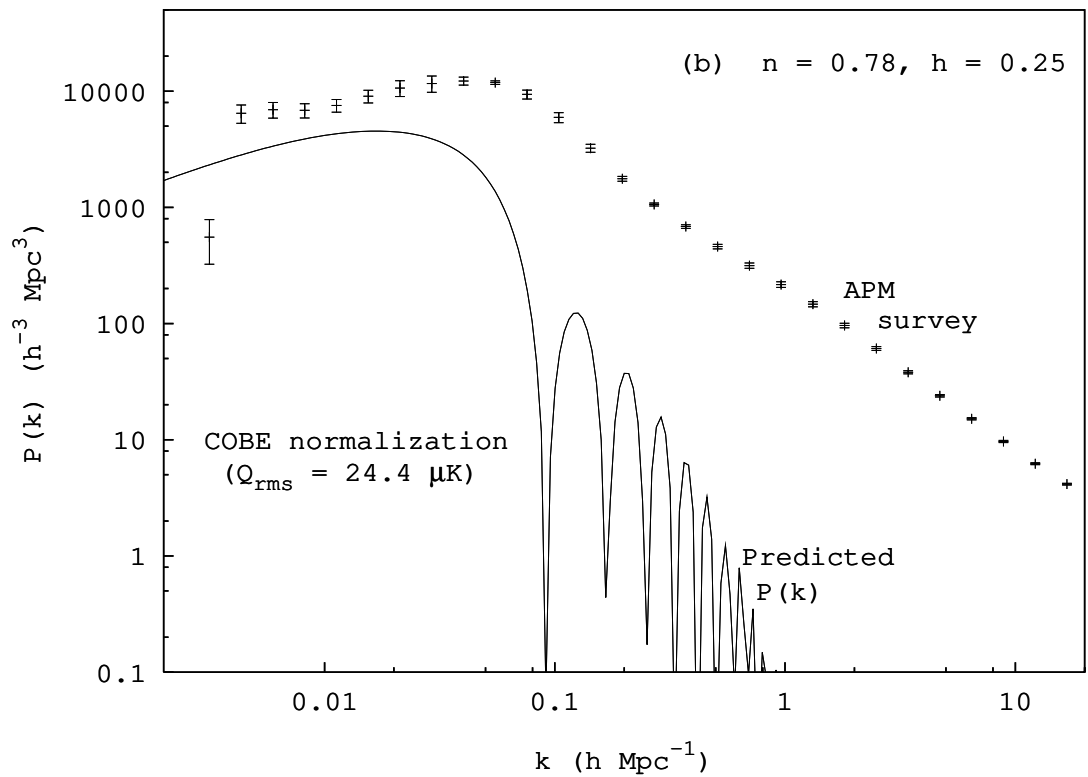}}
\medskip
\caption{Predicted power spectra of density fluctuations normalized to
 COBE for two $\Omega_{\rm B}=1$ models, (a) $n=0.5, h=0.62$ and (b)
 $n=0.78, h=0.25$, compared with APM data.}
\label{fig7}
\end{figure} 
\end{document}